\documentclass[a4paper,11pt,twocolumn]{article}
\usepackage[margin=1.5cm,top=.5cm]{geometry}
\usepackage{graphicx}
\usepackage[british,UKenglish,USenglish,english,american,russian]{babel}
\usepackage{amssymb}
\usepackage{amsmath}
\usepackage{cite}

\graphicspath{{figures/}}

\begin{document}
	\title{\textbf{The influence of local correlations on the phase states in the model of <<semi-hard-core>> bosons on a square lattice.}}
	
	\author{ Ulitko V. A. \and Konev V. V. \and Chikov A. A. \and Panov Y. D.}
	
	\date{Ural Federal University, 620002 Ekaterinburg, Russia}
	
	\maketitle
	
	\vspace{1em}

\begin{otherlanguage}{english}

\begin{abstract}
	
The work considers a model of charged \flqq semi-hard-core\frqq bosons on a square lattice with a possible occupation number $n = 0,1,2$ at each node. Temperature phase diagrams of the model are obtained using numerical Monte Carlo quantum simulation methods, and the influence of local charge correlations is examined. Comparison with results from mean-field methods shows that local charge correlations contribute to an increased role of quantum fluctuations in the formation of phase states. 
		
\end{abstract}
\end{otherlanguage}
	
\section{Introduction}

Low-dimensional  lattice boson models with competing interactions exhibit a rich set of quantum phase states and have been actively studied over the past few decades\cite{Heng2019,Hebert2001,Sengupta2005,Jiang2012,Schmidt2006,Capogrosso2008, Chen2008,Dutta2015}. 
Researchers are interested in such systems in connection with the experimental discovery of the competition between charge ordering and superconductivity in high-temperature superconductors \cite{Demler2004} and some bosonic phases, such as the Mott insulator phase and the superfluid phase for super-cold atoms in optical lattices \cite{ Greiner2002,Morsch2006,Zhou2009,Islam2015}. 
It should be noted that a long-lived bound state of two bosons in an optical lattice was discovered \cite{Winkler2006}, which makes it relevant to consider models with paired boson transfer, since it has ceased to be an object of purely theoretical constructions.

In this work, we consider the model of <<semi-hard-core>> bosons \cite{Dutta2015,Zhou2009}, which generalizes the Bose-Hubbard model on a square lattice \cite{Panov2018}.
The model is a system of charged bosons with possible filling at the node $n_i=0,1,2$. 
In this work, we do not take into account single-particle boson transport, since we decided to focus on systems with predominant two-particle transport. 
In the language of pseudospin description \cite{Batista2004}, the Hamiltonian of the system can be written using the pseudospin operator $S=1$ in the following form \cite{Moskvin2015,Panov2019}:

\begin{equation*}
	\label{Hfull}
	\begin{split}
		\hat{\cal H} = & \Delta \sum_i  \hat{S}_{iz}^2 + V \sum_{ <ij>} \hat{S}_{iz} \hat{S}_{jz} \\
		& -t \sum_{<ij>} (\hat{S}_{i+}^2 \hat{S}_{j-}^2 + \hat{S}_{i-}^2 \hat{S}_{j+}^2) - \mu \sum_i  \hat{S}_{iz},
	\end{split}
\end{equation*}

where the $z$-component of the pseudospin operator $\hat{S}_{iz}$ is related to the number of bosons at a site by the relation $\hat{n}_i = \hat{S}_{iz} + 1$. In this case, its eigenvalues $S_{iz} = \pm 1, 0$ correspond to the charge of bosons on a site, counted from the half-filling level, when $n_i = 1$ at all lattice sites. 
The first and second terms describe local and interstitial charge correlations, respectively.
The third term is responsible for the transfer of boson pairs between neighboring nodes.
The summation in the second and third terms is carried out over the nearest neighbors of the square lattice.
The last term is proportional to the chemical potential $\mu$ and allows us to take into account the condition of constancy of the total number of bosons $N$.
Using $x= n - 1$ - the deviation of the boson concentration from half filling, this condition can be written as follows:

\begin{equation}
	\label{xFix}
	x = \frac{1}{N} \sum_i S_{iz}
\end{equation}

Previously \cite{Panov2018} we studied the phase diagrams of the ground state of this model.
The qualitative difference between these diagrams in the $\Delta-x$ coordinates to the left and to the right of the Heisenberg point, defined by the condition $2V = t$, was demonstrated.
In this work we will restrict ourselves to the case of <<weak>> transfer: $2V > t$.
Then, depending on the relationship between the parameters of the Hamiltonian, the following phase states can be realized: charge ordering (CO) - an analogue of antiferromagnetic ordering along the $z$ axis, a superfluid phase (SF), and also, or a homogeneous mixture of these phases - - phase of a superfluid solid (supersolid, SS) or an inhomogeneous mixture of phases - phase separation (PS). 
In the work \cite{Micnas1984} the evolution of the temperature phase diagrams of this model in the atomic limit ($ t = 0 $) was studied.
It was shown that an increase in the local correlation parameter $\Delta$ also leads to a restructuring of the $T-x$ phase diagram of the system, in particular, to a shift in the maximum critical temperature $T_{CO}(x)$ for the CO phase. This motivated us to consider the influence of local correlations on temperature phase diagrams in the general case.
In the work \cite{Konev2021} the model of semi-rigid bosons was considered in the mean field method (MFA) approximation.
We obtained temperature phase diagrams and studied their evolution depending on the $\Delta$ parameter.
In the case of interest to us $2V > t$, we can distinguish two ranges of values for which the phase diagrams have qualitative differences. 
In the $\Delta \le 0$ limit, the model is similar to the case of “hard-core” bosons \cite{Robaszkiewicz1981}.
An increase in $\Delta$ makes states with $S_z = \pm 1$ unfavorable, the exchange between which generates two-boson transfer, and the SF phase region decreases.
This is accompanied by a decrease in the critical temperatures $T_{CO}(x)$ and $T_{SF}(x)$, as well as a shift in the CO-SF-PS tricritical point.
However, at $\Delta > \Delta^{*} = 0.75 t$ the situation changes.
The region of the SF phase continues to decrease, while the CO phase expands, and the $T_{CO}(x)$ dependence becomes nonmonotonic and, at sufficiently large $\Delta$, is described by a parabola with a maximum at the point $x = 0.5$.
In both regions, the system demonstrates a wide range of scenarios for the formation of phase states, such as order-to-order transitions, a change in the type of phase transition, and the appearance of new critical points.

Consequently, further study of the influence of local correlations on the phase states of this model is of interest 
In particular, the question arises about the influence of quantum fluctuations on the evolution of $T-x$ phase diagrams. 
In the work \cite{Konev2021} it was shown that for $\Delta \le 0$ the MFA results are in qualitative agreement with the results obtained by the quantum Monte Carlo method (QMC) \cite{Schmid2002}.
QMC allows us to more accurately take into account the off-diagonal correlations of the pseudospin operator associated with boson transport.
In this work, our main goal was to refine previous results using numerical simulations using the QMC method in the situation $\Delta > 0$.
In Section 2 we provide a description of the quantum algorithm we use.
Section 3 is devoted to the results obtained and their discussion.

\section{Quantum Monte-Carlo}
	
Like many other QMC algorithms, the SGF algorithm operates directly on physical states. An occupation number basis is chosen. We consider a Hamiltonian written in the form $\hat{H}=\hat{V}-\hat{T}$, where $\hat{V}$ is diagonal in the choosen basis, and $\hat{T}$ is off-diagonal and assumed to have positive matrix elements. The SGF algorithm does not require any further assumptions on the Hamiltonian. Deﬁning the inverse temperature $\beta$, the purpose of the algorithm is to sample the partition function $Z(\beta)=Tr \hspace{0.1cm} e^{-\beta \hat{V}} \hat{\sigma}(\beta),$ where $\hat{\sigma}(\beta)=T_{\tau} e^{\int^{\beta}_0 \hat{T}(\tau)d\tau}.$

To this end, we deﬁne the Green operator $\hat{G}$ by 

$$
\hat{G} = \sum_{p=0}^{\infty}\sum_{q=0}^{\infty}g(p,q)\sum_{[i_p,j_q]}
\prod_{k=1}^{p}\hat{S}_{i_k}^{+} \prod_{l=1}^{q} \hat{S}_{j_l}^{-}.
$$

and g is an arbitrary function, with the constraint g(0,0) = 1, that is all diagonal matrix elements of $\hat{G}$ are equal to 1. By breaking up the
exponential in statistical sum at imaginary time $\tau$ and introducing the Green operator between the two parts, we can deﬁne an extended partition function: 

$$
Z(\beta,\tau) = Tr e^{-(\beta-\tau) \hat{H} } \hat{G} e^{-\tau \hat{H} }.
$$

Deﬁning $\hat{T}(\tau) = e^{\tau \hat{V}} \hat{T} e^{-\tau \hat{V}}$ and $\hat{G}(\tau) = e^{\tau \hat{V}} \hat{G} e^{-\tau \hat{V}}$,
and using the equality

$$
e^{-\beta (\hat{V}-\hat{T})}=e^{-\beta \hat{V}} T_{\tau} e^{\int^{\beta}_0 \hat{T}(\tau)d\tau},
$$

where $T_{\tau}$ denotes the time-ordering operator over the variable $\tau$ with time increasing from the right to the left, the extended partition function takes the form:

$$
Z(\beta,\tau)=Tr \hspace{0.1cm} e^{-\beta \hat{V}} T_{\tau} \left[ e^{\int^{\beta}_{\tau} \hat{T}(\tau^{\prime})d\tau^{\prime}} \hat{G(\tau)} e^{ \int^{\tau}_0 \hat{T}(\tau^{\prime})d\tau^{\prime}} \right],
$$

By expanding the exponentials in  and ordering the operators in imaginary time, we get:

$$
Z(\beta) = e^{-\beta \hat{V}} \sum_{n \geqslant 0} \int_{0 < \tau_1 < \tau_2 < ... < \tau ... < \tau_n < \beta} \hat{T}(\tau_n) \hat{T}(\tau_{n-1})
... \hat{G(\tau)} ... \hat{T}(\tau_2) \hat{T}(\tau_1) d\tau_n ... d\tau_2 d\tau_1.
$$

By introducing complete sets of states between each $\hat{T}$ and$\hat{G}$ operators, and an extra set with for the trace, the extended partition function takes
the ﬁnal form

$$
Z(\beta, \tau) = e^{-\beta \hat{V}} \sum_{n \geqslant 0} \sum_{\Psi_1...\Psi_n}  \int_{0 < \tau_1 < \tau_2 < \tau_3 ... \tau_n < \beta}   \langle \Psi_0 \vert \hat{T}(\tau_n) \vert \Psi_{n-1} \rangle 
$$
$$... \langle \Psi_L \vert \hat{G}(\tau) \vert \Psi_R \rangle ... \langle \Psi_2 \vert \hat{T}(\tau_2) \vert \Psi_1 \rangle  \langle \Psi_1 \vert \hat{T}(\tau_1) \vert \Psi_0 \rangle d\tau_n ... d\tau_2 d\tau_1.
$$

The extended partition function Z($\beta$, $\tau$) is a sum of diagonal conﬁgurations that belongs to the actual partition function Z($\beta$), and non-diagonal conﬁgurations. More precisely,

$$
Z(\beta,\tau) = Z(\beta) +
$$
$$
+ \sum_{L \neq R}Tr \hspace{0.1cm} e^{-\beta \hat{V} } \hat{\sigma}(\beta, 
\tau)  
\vert L \rangle  \langle L \vert \hat{G}(\tau) \vert R \rangle  \langle R \vert
\hat{\sigma}(\tau,0).
$$

The purpose of the algorithm is to evolve from a diagonal conﬁguration to another one, via non-diagonal conﬁgurations. The role of the Green operator
$\hat{G}$ is to allow the transition from a conﬁguration to another one by propagating across the operator string and inserting or removing $\hat{T}$ operators while the oﬀset ﬂuctuates. By satisfying detailed balance, the conﬁgurations of the extended partition function can be generated with an extended Boltzmann weight. Then all quantities of interest can be estimated using those conﬁgurations.

\section{Influence of local correlations on phase states}

\begin{figure}[ht]
	\includegraphics[scale=0.55]{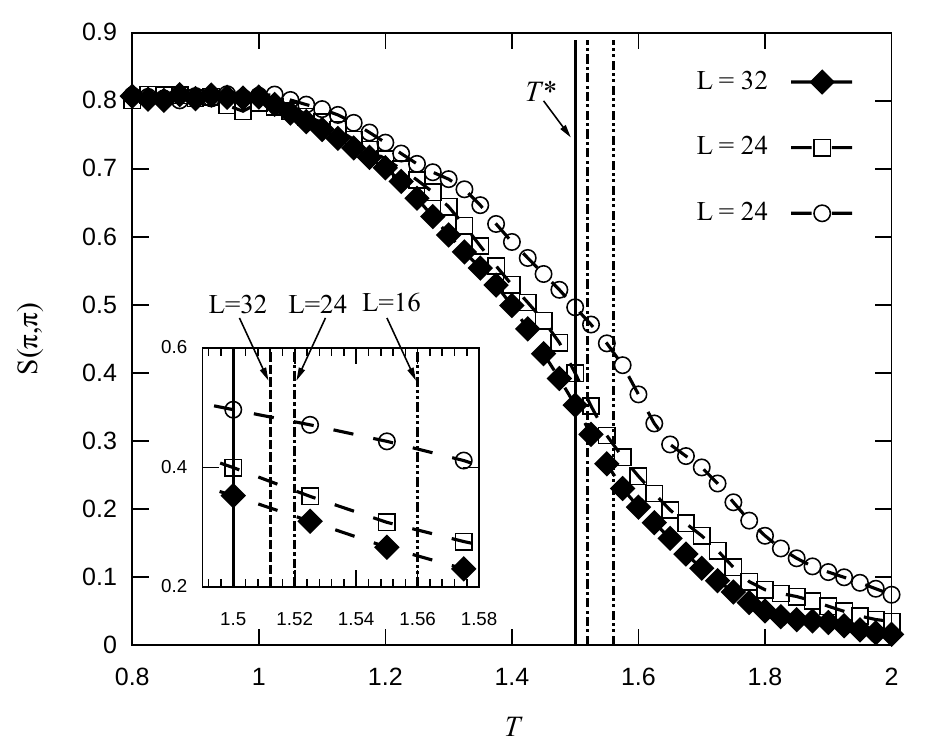} 
	\caption{\footnotesize{Temperature dependences of the structure factor $S_{CO}(\pi,\pi)$ for different lattice sizes $L$, obtained by the SGF method in the limit of <<hard>> bosons. Model parameters at $V/t = 0.75$, $\Delta / t = -2$, $x = 0$. The dotted vertical lines correspond to the obtained critical temperatures, and the value of $T^{*}$ is taken from \cite{Schmid2002}}}
	\label{fig:Shmidt}
\end{figure}

In view of the symmetry of the phase diagrams of the system relative to the point $x = 0$, calculations were performed for positive values of $x$.
In all MK simulations, we took $V/t = 0.75$, which corresponds to the parameter values from the work of \cite{Schmid2002}.
All temperatures are given in units of the transport integral $t$.
The simulation was carried out on a square lattice of size $L \times L$ with periodic boundary conditions.
As the order parameter of the CO phase, we used the structure factor of the average value of the operator $\hat{S_z}$ at the point $\vec{k}=(\pi,\pi):$
$$
S_{CO}(\vec{k})=\frac{1}{N^2} \sum_{i,j}^N  S_{iz}S_{jz} e^{-i\vec{k} \vec{r}_{ij}}, 
$$ 
The SF phase order parameter (stiffness) was calculated using the formula:
$$
\rho_s = (\langle W_x^2 \rangle + \langle W_y^2 \rangle)/2\beta t \,,
$$
where $\beta = 1/T$, and $W_x$ and $W_y$ are the winding numbers along the $x$ and $y$ directions on the lattice.
The PS phase in our case corresponds to the region where both order parameters are nonzero.
All critical temperatures were determined by the minimum derivative of the corresponding order parameter.
The SGF algorithm is extremely demanding on computing resources, so it was decided to limit the system size to $L = 24$.
The number of Monte Carlo steps was selected in such a way that the errors in the order and energy parameters did not exceed 5\%.
To test the results obtained, we carried out a series of simulations for various $L$ in the limit of <<hard-core>> bosons (Fig. \ref{fig:Shmidt}).


%
%
%

\begin{figure}[ht]
	\includegraphics[scale=0.7]{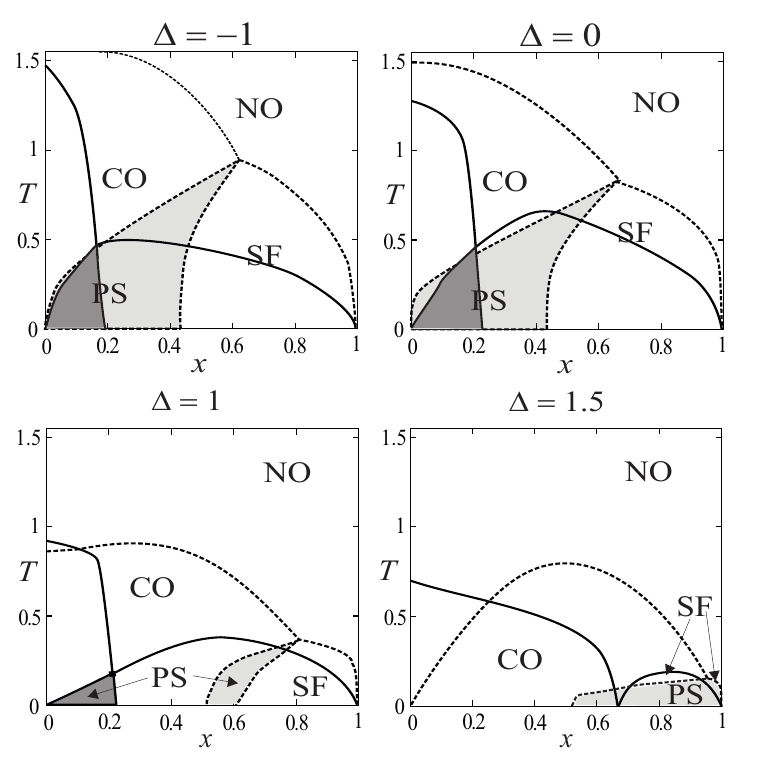} 
	\caption{\footnotesize{Temperature phase diagrams for the value $V / t = 0.75$. The dark (light) shaded area corresponds to the PS phase obtained by the SGF(MFA) method. The solid and dotted lines indicate the critical temperatures obtained by the SGF and MFA methods, respectively.}}
	\label{fig:PD}
\end{figure}

%
In Figure \ref{fig:PD}, a set of phase diagrams in the $T-x$ variables for a series of $\Delta$ values and $V=0.75$ obtained using the SGF and MFA methods is shown. The critical temperatures of the phases are indicated by solid (dashed) black lines, and the regions of phase separation are shown in dark gray (light gray) color for the SGF (MFA) method, respectively.

Comparing the phase diagrams obtained using the SGF and MFA methods allows us to understand the role of quantum fluctuations (non-local off-diagonal correlations) in the formation of the phase diagrams of the model. The temperature phase diagrams are significantly affected by quantum fluctuations.

It can be noted that for $\Delta=0$, a significant difference in the transition temperatures to the ordered phases is observed (see Figure \ref{fig:PD}).

However, as the local correlations increase, this difference becomes smaller. 
It is important to note that the phase boundaries in terms of filling fraction are significantly different. 
For instance, the region of CO phase in terms of filling fraction is much smaller in the SGF method, meaning that the quantitative ratio between the charge and the superfluid phases has shifted towards the latter compared to the MFA. 
Additionally, a shift in the value of local correlations at which significant changes occur in the phase diagrams can be observed. 
For the quantum Monte Carlo method, this value is around $\Delta_{SGF} \approx 1.1 t$, while for the mean field method it is $\Delta_{MFA} = 0.75 t$. 
This may be due to the correct accounting of quantum fluctuations in the SGF method, as this leads to higher energy values for the phases compared to the case of MFA.

\section{Conclusion}

Phase $T-x$ diagrams of the model were obtained using the SGF method, and their evolution was studied as the parameter of local correlations $\Delta$ increased. The general trend is the suppression of CO and SF phases with increasing $\Delta$. Comparison of these results with MFA data showed that the transition to a quantum algorithm is manifested in the renormalization of the $\Delta$ parameter. The SGF method qualitatively observes the same effects as MFA, but at higher values of $\Delta$. Additionally, the proportion of the SF phase on the diagram is higher in the SGF method than in MFA. Meanwhile, the position of the tricritical point in SGF is always at lower values of $x$. We explain this by the fact that the consideration of quantum fluctuations reduces the kinetic energy of the system, making the SF phase more favorable.
	
\bigskip
	
\textbf{Acknowledgement}
	
The research funding from the Ministry of Science and Higher Education of the Russian Federation (Ural Federal University Program of Development within the Priority-2030 Program) is gratefully acknowledged.

\end{document}